\documentclass[aps,prd,amsmath,showpacs,nofootinbib]{revtex4}
\RequirePackage[]{amssymb}
\usepackage{graphicx,euscript}

\def\({\left(}
\def\){\right)}

\newcommand{\beq}{\begin{equation}}
\newcommand{\eeq}{\end{equation}}
\newcommand{\bea}{\begin{eqnarray}}
\newcommand{\eea}{\end{eqnarray}}

\newcommand{\bean}{\begin{eqnarray*}}
\newcommand{\eean}{\end{eqnarray*}}
\newcommand{\bs}{\begin{subequations}}
\newcommand{\es}{\end{subequations}}

\DeclareMathOperator{\const}{const}
\newcommand{\arcosh}{\textrm{arcosh}}
\newcommand{\arctanh}{\textrm{arctanh}}
\newcommand{\sech}{\textrm{sech}}

\begin{document}

\title{On the cylindrically symmetric  wormholes  $\rm WhCR^e$: The motion of  test particles.}

\author{Asya V. Aminova}
\email{asya.aminova@kpfu.ru}
\affiliation{Department of General Relativity and Gravitation, 
 Kazan
Federal University,  18 Kremlyovskaya St., Kazan 420008,
Russia}

\author{Dieter R. Brill}
\email{brill@umd.edu}%
\affiliation{Maryland Center for Fundamental Physics, University of Maryland, College Park, MD 20742, USA
}%

\author{Pavel I. Chumarov}%
\email{p.i.t.choumarov@mail.ru}%
\affiliation{Department of General Relativity and Gravitation, 
Kazan
Federal University,  18 Kremlyovskaya St., Kazan 420008,
Russia} %

\author{Aleksandr Yu. Shemakhin}%
\email{Aleksandr.Shemakhin@kpfu.ru}%
\affiliation{Department of Radiophysics, Institute of Physics, Department of Mathematical Statistics, Institute of Computer Mathematics and Information Technologies, Kazan
Federal University,  18 Kremlyovskaya St., Kazan 420008,
Russia}

\begin{abstract}

 In this article we partially implement the program outlined in the previous paper of the authors \cite{AmChum2013}. The program owes its origins to
  the following comment
 in paper \cite{CvKK}, where a class of static spherically symmetric solutions in $(4+n)$-dimensional Kaluza--Klein theory  was studied: ``...We suspect that the same thing [as for spherical symmetry] will happen for axially symmetric stationary configurations, but it remains to be proven".
  We study the  radial and non-radial motion of  test particles in the cylindrically symmetric wormholes  found in \cite{AmChum2013} of  type $\rm WhCR^e$ in 6-dimensional reduced Kaluza--Klein theory with Abelian gauge field and two dilaton fields, with particular attention to the extent to which the wormhole is traversable. In the case of non-radial motion along a hypersurface $z=\const$ (``planar orbits")
 we show  that, as in the Kerr and Schwarzschild
 geometries  \cite{chan},  we should distinguish between orbits with  impact parameters greater resp. less than a certain critical value $D_c$, which corresponds to the unstable  circular orbit of radius $u_c$ $(r_c)$.\footnote{Note the difference in  the notations: in \cite{chan} $u=1/r$ where $r$ is the radial spherical coordinate, here $u=\ln r$ where $r$ is the radial cylindrical coordinate.} For $D^2>D_c^2$ there are two kinds of orbits: orbits of the first kind  arrive from infinity and have pericenter distances greater than $u_c$, whereas  orbits of the second kind have apocenter distances less than $u_c$ and terminate at the singularity at $u=-\infty$ $(r=0)$. For $D=D_c$ orbits
 of the first and second  kinds merge and both orbits spiral an infinite number of times toward the unstable circular orbit  $u=u_c$.
 For $D^2<D_c^2$ we have only  orbits
 of one kind:  starting at infinity, they cross  the wormhole throat and terminate at the
 singularity.

\end{abstract}


 \maketitle

 \section{Introduction.}
  The general static cylindrically symmetric  space-time metric can be written in the form
 \begin{equation}\label{cm}
 ds ^ 2 = e ^ {2 \gamma (u)} dt ^ 2 - e ^ {2 [\beta (u) +\gamma (u) +\xi (u)] } du ^ 2 - e ^ {2 \xi (u)} dz ^ 2 - e ^ {2 \beta (u)} d \phi ^ 2,
 \end{equation}
 where $ u \in (- \infty, + \infty)$ is a cylindrical
radial  coordinate, $ z \in (- \infty, + \infty) $ is the longitudinal
coordinate, and
  $ \phi \in [0, 2 \pi] $ is the angular coordinate (see \cite{1}).

 The  metric~\eqref{cm} has
one timelike Killing vector $ {\xi} _1 =
\partial_ t $ and  two spacelike Killing vectors $  {\xi} _2 =
\partial_ \phi $,
$  {\xi} _3 = \partial_z $, which define the axial symmetry. It is invariant under the simultaneous interchange $z \leftrightarrow \phi$ and $\xi \leftrightarrow \beta$; one of the coordinates ($\phi$) was arbitrarily chosen to be periodic with period $2\pi$ to represent a cylindrically symmetric (rather than plane symmetric) geometry. (Frequently the scale of $\phi$ is chosen uniquely to make the origin nonsingular, but this cannot be done for the metrics considered here.)

 Radial null geodesics are described by
 $du/dt = \pm e^{-(\beta(u) + \xi(u))}$, which will be non-zero for all $u$.
 Therefore there is no horizon, and any singularity at the origin is visible from everywhere (``naked").

By definition, metric~\eqref{cm} describes {\it a cylindrically symmetric wormhole} $V$
if the ``circle
radius''  $\rho(u): = e ^ {\beta (u)}$
has an absolute minimum  $\rho (u_0) > 0$ at some point $u = u_0$ and for all
possible values of $u$ the metric functions  $\beta(u)$,
$\gamma(u)$, and $\xi(u)$  are smooth and finite
\cite{1}. %

 A cylindrical
hypersurface $\Sigma_0$ defined by the equation
$
u = u_0
$
is called {\it a throat}  of the
wormhole $V$, if $V$
can be presented as the  union
$V=V_{-}\cup \Sigma_{0}\cup V_{+}$, where  $V_{-}=\{(t,\ u,\ z,\ \phi)\in V |\ u< u_{0}\}$ and
 $V_{+}=\{(t,\ u,\ z,\ \phi)\in V |\ u> u_{0}\}$ are interpreted as the two universes at the ends of the wormhole.

  The  wormhole is called
{\it traversable} if travel is possible from one universe ($V_{-}$ or $V_{+}$) to another ($V_{+}$ or $V_{-}$, respectively), i.e.,  if
    a traversing   timelike path
through the wormhole's throat is allowed in a finite
time (see, for example,~\cite{3}).

In~\cite{AmChum2013}
   we studied cylindrically symmetric Abelian wormholes in $(4 +n)$-dimensional Kaluza-Klein theory. It was
shown that static four-dimensional cylindrically symmetric solutions in $(4 + n)$-dimensional Kaluza-
Klein theory with maximal Abelian isometry group $U(1)^n$ of the internal space with diagonal internal
metric can be obtained (as in the case of a supersymmetric static black hole \cite{CvYou}) only if the isometry group of the internal space is broken down to the
$U (1)_e \times U(1)_m$ gauge group; the solutions correspond to dyonic configurations with one electric $(Q_e)$ and one
magnetic $(Q_m)$ charge that are related either to the same $U (1)_e $ or $ U(1)_m$ gauge field or to different factors
of the $U (1)_e \times U(1)_m$ gauge group of the effective six-dimensional Kaluza-Klein theory. We found new static cylindrically symmetric
exact solutions of the six-dimensional Kaluza-Klein theory with two Abelian gauge fields
$A_\mu$, $B_\mu$, a dilaton field $\psi$, and a scalar field $\chi$
 associated with the internal metric. We obtained new types of static cylindrically symmetric
wormholes supported by  radial and longitudinal electric and magnetic fields.
 In the case of radial gauge fields we found three  types of static cylindrically symmetric wormholes:  dyonic ${\rm WhCR^{e;m}}$ with nonzero electric and magnetic charges,
$\rm WhCR^e$ with nonzero electric charge,
and  $\rm WhCR^m$ with nonzero magnetic charge.
For longitudinal gauge fields we obtained nine types of  dyonic wormholes
${\rm WhCL}^{k|e;j|m}$,  $k,j = 1, 2,3$, the wormhole  ${\rm WhCL}^{3|e}$ with nonzero electric charge, and   the wormhole ${\rm WhCL}^{3|m}$ with nonzero magnetic charge.
From the physical point of view these wormholes are interesting because they do not need exotic matter or phantom fields for support. However the universes $V_-$ and $V_+$ at the ends of the wormhole are not benign, for example $V_-$ contains a curvature singularity. We will leave exploration of these features to a later investigation and focus here primarily on the interesting, near-throat region.

In this article we consider  the geodesic structure of the wormhole solutions of  type ${\rm WhCR^e}$ with the space-time metric in coordinates $t,r=e^u,z,\phi$ (\cite{AmChum2013},  Eqs. (35)-(38))
 $$ ds^2 =  k_e\Lambda_e \left( \frac{r^{2c}dt^2}{k_e^2\Lambda_e^2}
 - r^{2(a+b-c-1)} d r^2 -  r^{2(b-c)} d {z}^2  -  r^{2(a-c)}
  d {\phi}^2\right)\quad (a,b,c
= \mathrm{const}),
  $$
 where
 $$
k_e=\sqrt{|q_e|}\equiv\sqrt{|Q_e/h_e|}, \quad
\Lambda_e=\sqrt{(r/r_e)^{h_e} +
(r/r_e)^{-h_e}},\quad 4ab=h_e^2+16c^2 $$$$
(h_e, q_e  = \mathrm{const},h_eq_e\neq 0).
 $$
  The throat radius  of the $\rm WhCR^e $ is
 $$
  r_0=r_e\left(\frac{h_e-4(a-c)}{h_e+4(a-c)}\right)^{1/(2h_e)}, \quad h_e>4|a-c|>0,
 $$
and the wormhole
 is generated by  the following   Abelian gauge field $A_\mu$,
 dilaton  $\psi$, and scalar field  $\chi$:
 $$
\begin{array}{c}
A_ \mu = \left((-1/(4q_e))\left[(r/r_e)^{h_e} -
(r/r_e)^{-h_e}\right]\left[(r/r_e)^{h_e} +
(r/r_e)^{-h_e}\right]^{-1},\ 0,\ 0,\ 0\right), \\ \\
 \psi =(1/(2\sqrt{2}))\ln\left(|q_e|
 r^{4c}[(r/r_e)^{h_e} +
(r/r_e)^{-h_e}]\right),\\ \\
 \chi=-(1/(4\sqrt{2}))\ln\left(|q_e|
 r^{-4c}[(r/r_e)^{h_e} +
(r/r_e)^{-h_e}]\right).\end{array} $$

  Further in this paper we put $r_e=1, c=0, h_e\equiv h$ and consider a 3-parameter family of the wormholes  $\rm WhCR^e$ with the line interval
 \begin{equation}\label{CWH_2}
 ds^2 =  \frac{dt^2}{\sqrt{|q_e|}\sqrt{r^{h} +
r^{-h}}}
 -  \sqrt{|q_e|}\sqrt{r^{h} +
r^{-h}}\left(r^{2(a+b-1)} d r^2 +  r^{2b} d {z}^2  +  r^{2a}
  d {\phi}^2\right)
  \end{equation}
  and the throat radius
  $$r_{0}=\left(\frac{1-4a/h}{1+4a/h}\right)^{1/(2h)}$$
  defined in coordinates $t,u,z,\phi$ by the formulas:
   \begin{equation}\begin{array}{c}
   \label{cm1}\begin{displaystyle}
    ds^2 =  \frac{dt^2}{\sqrt{2|q_e|\cosh (h u)}}
 -  \sqrt{2|q_e|\cosh (h u)}\left(e^{2(a+b)u} d u^2 +  e^{2bu} d {z}^2  +  e^{2au}
  d {\phi}^2\right), \end{displaystyle} \\ \\
   4ab=h^2>0, \ \ 4|a|<h<b, \ \
q_e\neq 0.
\end{array}\end{equation}
Further, we put $a>0$, whence $b=h^2/(4a)>0$,  $ 4a/h <1$, and $r_0<1$,  since the opposite choice $ a=-\alpha^2<0, b=-h^2(4\alpha^2)$ is reduced to the previous one by a change $r\to 1/r$. Metric~\eqref{CWH_2} and its geodesics
as well as dilaton field $\psi$ and scalar field $\chi$ are invariant under this change, while the gauge field changes sign.

Metric~\eqref{CWH_2} can be obtained  from \eqref{cm} by putting
\begin{equation}\label{metr(2)}
\beta =au+\frac{1}{4}\ln\left( 2|q_e|\cosh (h u)\right),\  \gamma =- \frac{1}{4}\ln\left(2|q_e|\cosh (h u)\right),\ \xi =bu+ \frac{1}{4}\ln\left( 2|q_e|\cosh (h u)\right).\end{equation}

 The hypersurface
 \begin{equation}\label{u0}
 u=u_{0} \equiv -h^{-1} \arctanh(4a/h)
\end{equation}
is the throat of wormhole~\eqref{cm1},   which is
 generated by  the following vector, dilaton, and scalar fields:
 $$
 \begin{displaystyle}
A_ \mu = \left(-\frac{1}{4q_e}\tanh (hu),\ 0,\ 0,\ 0\right), \quad
 \psi =-2\chi=\frac{1}{2\sqrt{2}}\ln\left(2|q_e|
\cosh (hu)
 \right).\end{displaystyle}
$$
  Note that metric~\eqref{cm1} is not asymptotically flat, which confirms the ``no-go" statement  about
the nonflat asymptotic behavior of a cylindrically symmetric
wormhole  in the absence of ghost fields, i.e.,  fields having negative kinetic energy~\cite{1}.

As the main source of information about the structure of any physical field is the behavior of test bodies, we will focus on the motion of test particles in the wormhole $\rm WhCR^e$ whose trajectories are geodesics.

  The geodesic motion in space--times of spherically symmetric
wormholes~\cite{Visser} was
studied, for example, in~\cite{3},~\cite{ttravel}.

 In~\cite{3} a detailed derivation  of  solutions of the Einstein field equations was presented, which describe traversable wormholes that, in principle, could be traversed by human beings.

  The creation of wormholes and their conversion into time machines   as well as  quantum-field-theoretic stress-energy tensors that are required to maintain a two-way traversable wormhole were   discussed in~\cite{ttravel}.

In~\cite{tog1}  the
motion of massive and massless test particles in a space-time of a slowly rotating spherically symmetric wormhole  with a ghost scalar field  as a source was considered, and  it was shown that  after crossing the wormhole throat the particles (massive or massless) moving  initially radially will move in a spiral   away  from the throat.

 The study of  dynamics of null and timelike geodesics  for  traversable static spherically symmetric Schwarzschild and Kerr thin--shell wormholes constructed by cut-and-paste method was presented  in~\cite{kag}.

 The radial geodesic motion of a massive particle into a version of an  Einstein--Rosen bridge was considered in~\cite{Poplawski}. This wormhole was constructed by gluing regions I and III of the Kruskal space-time along the future resp. past horizons, which requires a delta-function matter source violating the energy conditions. This wormhole is traversable by complete timelike geodesics. The author suggests  that observed astrophysical black holes may be Einstein--Rosen bridges, each with a new universe inside.

  In~\cite{Hackmann} the geodesic motion of  charged test particles  in the gravitational field of  a rotating and electromagnetically charged Kerr--Newman black hole was studied;
 the colatitudinal and radial motions of particles moving along timelike world lines were classified.

   A class of axially symmetric stationary exact solutions of the phantom scalar field in general relativity  describing rotating and magnetised wormholes was found in~\cite{Matos2}.

   In ~\cite{Matos} properties of a Kerr-like wormhole supported by phantom matter were studied.
   The geodesics of the Kerr-like phantom wormhole  were analysed   in~\cite{Miranda}, where it was shown that the wormhole can be traversable for an observer like a human being.

    This article is organized as follows. In section II we discuss the general properties of geodesic motion in space-time~\eqref{cm1}.
We compute Kretschmann's invariant $K= R_{\mu\nu\lambda\rho} R^{\mu\nu\lambda\rho}$ and show
that the space--time~\eqref{cm1} becomes singular as  $u\to -\infty $
 and has a physical (naked) singularity at $u = -\infty$ (for $a,b >0$).
In section III we describe radial motion of massive test particles.
   We find the turning points
   $u_\pm$ of the particle motion and
the regions accessible by a particle with  energy $E$.
Equating $u_-$ and $u_+$, we find $u=0$, $E=1/\sqrt[4]{2|q_e|}\equiv E_0.$ We prove that the
particle located at $u=u_-=u_+=0$  will be at the point of unstable equilibrium and study in detail  the character of motion of particles with energies $E$ greater and/or less than $E_0$.
We find   a (lower)  energy threshold $E_{thr}\equiv E_0\sqrt[8]{1- 16a^2/h^2}$ of  traversability of the wormhole throat for a radially moving massive test particle.
In section IV we consider
radial trajectories of photons and prove that  each of such trajectories  crosses the throat.
Non-radial motion along a hypersurface $z=\const$ (``planar orbits") is studied in section V where
 we draw parallels with the Kerr and Schwarzschild %
geometries.
In section VI we investigate
non-radial motion in the plane $\phi=\mbox{const}$. We prove that all null orbits have a pericenter distance  $u_1$ and terminate at radial infinity  and identify  four types of massive particle behavior. Conclusions are given in section VII.

 \section{The general properties of geodesic motion in the space-time~(1).}\label{II}

 According to the general theory of relativity,  trajectories $x^\mu = x^\mu (\tau)$ of  test particles in the  absence of force fields are  geodesics, which are  integral curves of the
 equations
 \begin{equation}\label{geo}
 \frac{Dv^\mu}{d\tau}\equiv \ddot x^\mu + \Gamma^\mu_{\nu\sigma} \dot x^\nu \dot x^\sigma=0,
 \end{equation}
 where $  {D} /{d \tau} $ denotes the absolute derivative with respect to an affine parameter $ \tau $, $ v ^ \mu :=dx^\mu /d\tau $ is the 4-velocity vector of the test particle, $\Gamma^\mu_{\nu\sigma}$ are the Christoffel symbols, and overdots denote derivatives with respect to  $ \tau $.
 For metric~\eqref{cm} the non-vanishing Christoffel symbols are
 $$\begin{array}{c}
\Gamma^t_{tu}=\gamma',\ \  \Gamma^u_{tt}=e^{-2(\beta+\xi)}\gamma',\
\ \Gamma^u_{uu}=\beta'+\gamma'+\xi',\\ \\  \Gamma^u_{zz}=-e^{-2(\beta+\gamma)}\xi', \ \ \Gamma^u_{\phi \phi}=-e^{-2(\gamma+\xi)}\beta', \ \ \Gamma^z_{uz}=\xi', \ \ \Gamma^\phi_{\phi u}=\beta',
 \end{array}$$
 and Eq.~\eqref{geo} gives
 \begin{equation}\label{geo1110}
  \dot v^t + 2\gamma' v^t v^u =0,
\end{equation}
\begin{equation}\label{geo1111}
 \dot v^u + (\beta'+\gamma'+\xi') (v^u)^2 + \gamma' e^{-2( \beta+\xi)}(v^t)^2 -
\xi' e^{-2( \beta+\gamma)}(v^z)^2
- \beta' e^{-2(\gamma+\xi)}(v^\phi)^2
 =0,
\end{equation}
 \begin{equation}\label{geo1112}
  \dot v^z + 2\xi' v^z v^u =0,
\end{equation}
 \begin{equation}\label{geo1113}
\dot v^\phi + 2\beta' v^\phi v^u  =0
\end{equation}
 ($v^t\equiv \dot t, \ v^u\equiv \dot u, \ v^z\equiv \dot z, \ v^\phi \equiv \dot \phi$, and the  prime denotes the derivative with respect to the radial coordinate $u$).

 The  equations of motion    can also be obtained from the Lagrangian
 $$
\EuScript{L}\displaystyle
 = \frac{1}{2} g_ {\mu \nu}  {\dot x} ^ \mu   {\dot x} ^ \nu
  \equiv\frac{1}{2}  \left( e^{2\gamma} \dot{t}^2 - e^{2(\beta+\gamma+\xi)} \dot{u}^2 - e^{2\xi} \dot{z}^2 -    e^{2\beta} \dot{\phi}^2 \right).
$$
 The  canonical momenta are
\begin{gather}
p_t \equiv \frac{\partial \EuScript{L}
}{\partial\dot{t}} =e^{2\gamma} \dot t,\quad
p_u \equiv \frac{\partial \EuScript{L}
}{\partial\dot{u}} = -e^{2(\beta+\gamma +\xi)}\dot u, \notag\\
p_z \equiv \frac{\partial \EuScript{L}
}{\partial\dot{z}} =-e^{2\xi}\dot z,\quad
p_\phi \equiv \frac{\partial \EuScript{L}
}{\partial\dot{\phi}} =-e^{2\beta} \dot \phi,
\end{gather}
and the Hamiltonian function is $ H
= v ^ \mu p_ \mu - \EuScript{L}
 = \EuScript{L}.$
From the Hamilton's equations $ \dot {p} _ \mu = -  \partial H/\partial x^\mu
$ we find
$$
 \dot{p}_t = - \frac{\partial   \EuScript{L}
 }{\partial t} =0, \quad  \dot{p}_z = - \frac{\partial   \EuScript{L}
 }{\partial z} =0, \quad
 \dot{p}_\phi = - \frac{\partial   \EuScript{L}
 }{\partial \phi} =0.
$$
By integrating, we obtain
first
integrals of the geodesic equations~\eqref{geo}
in the cylindrically
symmetric space-time~\eqref{cm}:
\begin{equation}\label{iiint1}
p_t=e^{2\gamma}\dot t\equiv E=\const >0,\qquad
p_z=-
e^{2\xi}\dot z\equiv - M_z=\const, \qquad
p_\phi=-
e^{2\beta}\dot\phi\equiv - L=\const.
\end{equation}
They are generated by the three Killing vector fields $\partial /\partial t, \ \partial /\partial z$, and $\partial /\partial \phi$. The constants of motion $E$ and $L$ are interpreted as the energy and angular momenta per unit mass of the particle with non-zero rest mass.

From Eq. \eqref{iiint1} we have
\begin{equation}\label{dtdzdphi}
t(\tau)  =t_0+E \int_{\tau_0}^\tau e^{-2\gamma(u(\bar\tau))} d\bar\tau, \quad z(\tau)  =z_0+
M_z \int_{\tau_0}^\tau e^{-2\xi(u(\bar\tau))} d\bar\tau, \quad \phi(\tau)  =\phi_0+
L \int_{\tau_0}^\tau e^{-2\beta(u(\bar\tau))} d\bar\tau,
\end{equation}
here $\tau_0$ denotes the initial value of the  parameter $\tau$ and  the integration constants $t_0$, $z_0$, and $\phi_0$  are the initial values of the coordinates $t$, $z$, and $\phi$, respectively.

From Eq.~\eqref{iiint1} and the equality $ v_ \mu v ^ \mu = \iota $, where   $ \iota =1$   for
time-like geodesics and $ \iota =0$  for null geodesics, we get
\begin{equation}\label{skorparticle}
\dot {u} ^ 2=(E^2e^{-2\gamma(u)}-M_z^2e^{-2\xi(u)}-L^2e^{-2\beta(u)}-\iota)  \exp[-2(\beta(u)+\gamma(u)+\xi(u))]\equiv f(u).
\end{equation}
We write Eq.~\eqref{skorparticle} in the form of an energy conservation law~\cite{chan}
$$
\dot {u} ^ 2   e^{2[\beta(u)+2\gamma(u)+\xi(u)]}  +\iota e^{2\gamma(u)} + M_z^2e^{2(\gamma(u)-\xi(u))}+L^2e^{2(\gamma(u)-\beta(u))}=E^2,
$$
or, in terms of an effective potential,
$$
\dot {u} ^ 2   e^{2[\beta(u)+2\gamma(u)+\xi(u)]}  + V_{eff}^2(u)=E^2,
$$
  where $  E  $ is the conserved
energy and
\begin{equation}\label{potencialnaya_energiya}
 V_{eff}(u) =\sqrt{ \iota e^{2\gamma(u)} + M_z^2e^{2(\gamma(u)-\xi(u))}+L^2e^{2(\gamma(u)-\beta(u))}}
\end{equation}
is  the effective potential for the geodesic motion.

By virtue of
~\eqref{dtdzdphi}
 Eqs.~\eqref{geo1110}, \eqref{geo1112}, and \eqref{geo1113} are satisfied identically, and Eq.~\eqref{geo1111}  is a differential consequence of Eq.~\eqref{skorparticle}.
We rewrite Eq.~\eqref{geo1111} in   the form
\begin{equation}\label{uskorenie_dlya_metriki_cm1}
\ddot u = -(\gamma'+\beta'+\xi')f(u)- E^2\gamma' e^{-2\gamma(u)}+M_z^2\xi'e^{-2\xi(u)}+L^2\beta'e^{-2\beta(u)} \exp[-2(\beta(u)+\gamma(u)+\xi(u))].
\end{equation}

By integrating ~\eqref{skorparticle} and applying~\eqref{dtdzdphi}, we get functions $t=t(\tau)$, $u=u(\tau)$, $z=z(\tau)$, and $\phi=\phi(\tau)$ describing the classical (non--quantum) motion of  uncharged point particle.

In the following sections  we consider
three possible cases: 1) $M_z=L=0$, 2) $M_z=0$,  $L\neq 0$, and   3) $L= 0$,  $M_z\neq 0.$

Non-zero components of Riemann curvature tensor $R_{\mu\nu\lambda\rho}$ of metric~\eqref{cm1}  are
$$\begin{array}{l}
R_{u\phi u\phi} =-(h/4)  \tanh(h u) (2|q_e|\cosh (h u))^{1/2}  [h\tanh(h u) +b ]e^{2au}, \\ \\
R_{uz uz} = -( h/4) \tanh(h u) (2|q_e|\cosh (h u))^{1/2} [h\tanh(h u) +a ] e^{2bu}, \\ \\
R_{tu tu} = -( h/4)  (2|q_e|\cosh (h u))^{-1/2}[ (3 h/2)  \tanh^2(h u) + (a+b)\tanh(h u)-h], \\ \\
R_{z\phi z\phi} = ( h/4) (2|q_e|\cosh (h u))^{1/2} [( h/4)  \tanh^2(h u) + (a+b)\tanh(h u) +h ], \\ \\
R_{t\phi t\phi} = ( h/4) (2|q_e|\cosh (h u))^{-1/2} [ ( h/4)  \tanh^2(h u) + a\tanh(h u) ]e^{-2bu}, \\ \\
R_{tz tz} = ( h/4) (2|q_e|\cosh (h u))^{-1/2} [ ( h/4) \tanh^2(h u) + b\tanh(h u) ]e^{-2au} .
\end{array}
$$
From here we can calculate  the Ricci tensor $R_\mu^\nu = g^{\nu\rho} R^\lambda_{\ldotp \rho \lambda \mu} $
 and scalar curvature $R=R_\mu^\mu$:
$$\begin{array}{l}
 R_t ^ t = -  (h^2/4) (2|q_e|)^{-1/2}(\sech (h u))^{5/2} e^{-2(a+b)u}, \\ \\
 R_u ^ u = - (h^2/8)  (2|q_e|\sech (h u))^{1/2} [3\tanh^2(h u) + 2(\sech (h u))^{2} ]e^{-2(a+b)u}, \\ \\
 R_z ^ z = (h^2/4)  (2|q_e|)^{-1/2} (\sech (h u))^{5/2} e^{-2(a+b)u}, \\ \\
 R_\phi ^ \phi =(h^2/4)  (2|q_e|)^{-1/2} (\sech (h u))^{5/2} e^{-2(a+b)u}, \\ \\
 R=-(3h^2/8)  (2|q_e|\sech (h u))^{1/2} \tanh^2(h u) e^{-2(a+b)u},
 \end{array}
$$
and from these we can compute the Kretschmann invariant $K= R_{\mu\nu\lambda\rho} R^{\mu\nu\lambda\rho}$:

$$
\begin{array}{c}
  K = (1/4) |q_e|^{-5}  [ c_1 + c_2\sech(h u) \sinh(h u)
-c_3 \sech^{2}(h u)- \\ \\ - c_4 \sech^{3}(h u) \sinh(h u)+71 h^4\sech^{4}(h u)]\sech(h u)e^{-4 (a+b) u}, \end{array}
$$
where $
c_1= 24 a^2 b^2+ (2a^2 + 3 a b +2b^2)h^2+(7/32) h^4,\
c_2=(a+b)(8ab+h^2)h, \
c_3= [2a^2+  11 a b +2 b^2+(15/16) h^2]h^2$, and $
c_4=3 (a + b)h^3$.

It can be easily verified that  $ \lim_{u\rightarrow -\infty} K = +\infty$ (for $a,b >0$).
From here it follows that space--time~\eqref{cm1} becomes singular as  $r\to 0$
 and has a physical  singularity at $u = -\infty \ (r=0)$ (for $a,b >0$).

\section{Radial motion of massive test particles.}\label{4.1.1}

We consider the first case: $ M_z = L = 0 $. From~\eqref{iiint1} we have $ z = \const $ and $ \phi = \const $. Hence, only $t$ and $u$ can depend on $\tau$.   The motion in this case is called radial.

From Eqs.~\eqref{metr(2)},~\eqref{dtdzdphi}--\eqref{uskorenie_dlya_metriki_cm1} one can obtain the following equations
for  time-like geodesics:
$$
  t = t_0+E\sqrt{2|q_e|}  \int_{\tau_1}^{\tau_2} \sqrt{\cosh(h u(\bar\tau))}d\bar\tau ,
   \ \ \ z(\tau)=z_0, \ \ \ \phi(\tau)=\phi_0,$$
 \begin{equation}\label{rad_skor}
 \dot {u} ^ 2 =  P_{rm}(E,u)  e^{-2(a+h^{2}/(4a))u},
\end{equation}
\begin{equation}\label{uskorenie}
 \ddot u = S_{rm}(E,u)e^{-2(a+h^{2}/(4a))u},
\end{equation}
$$
 V_{eff}(u) = \frac{1}{\sqrt[4]{2|q_e|\cosh(h u)}},$$
where we have denoted
\begin{equation*}P_{rm}(E,u)\equiv
E^2-\frac{1}{\sqrt{2|q_e|\cosh(hu)}},\
 S_{rm}(E,u)\equiv  \frac{h}{4}\frac{\tanh(hu)}{\sqrt{2|q_e|\cosh(hu)}} - (a+h^{2}/(4a)) P_{rm}(E,u).
\end{equation*}
The plot of the effective potential
$V_{eff}(u)$ for radial timelike geodesics in space-time~\eqref{cm1} at fixed $h =1$, $q_e =0.5$, and $ a=0.125$   is shown   in Fig.~\ref{V_{eff}(u)}.

\begin{center}
\begin{figure}[h]
\center{\includegraphics[width=0.6\linewidth]{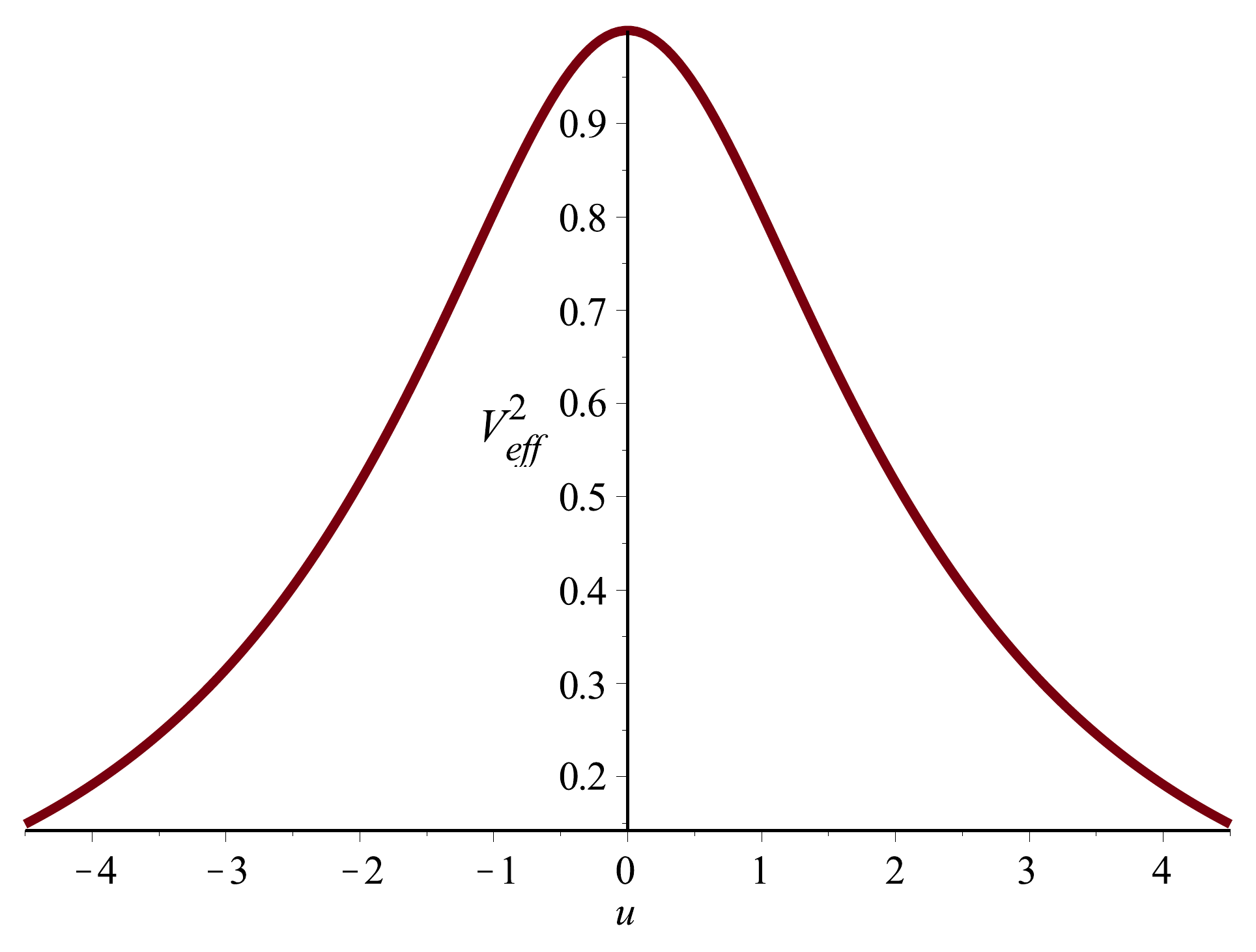}}
\caption{The plot of the squared effective potential
$V^2_{eff}(u)$ for radial timelike geodesics ($M_z=L=0$) in space-time~\eqref{cm1} at fixed $h =1$, $q_e =0.5$, and $ a=0.125$.}
  \label{V_{eff}(u)}
\end{figure}
\end{center}

It follows from Eq.~\eqref{rad_skor}  that the function $P_{rm}(E,u)$ must be nonnegative   \begin{equation*}
P_{rm}(E,u)= E^2-\frac{1}{\sqrt{2|q_e|\cosh(hu)}}\geq 0.\end{equation*}
This condition is satisfied for all real $
     u\in (-\infty, +\infty) $ when \ $E \geq 1/\sqrt[4]{2|q_e|}$  and for
$ u \in (-\infty,u_-] \cup  [u_+, +\infty)$ with
  \begin{equation}\label{reg}  u_\pm=\pm h^{-1} \mid\arcosh[(2|q_e|E^4)^{-1}]\mid \end{equation}
   when \ $E \leq 1/\sqrt[4]{2|q_e|}$.

   The turning points
   of the particle motion are defined by equation $\dot u=0$ or, by virtue of~\eqref{rad_skor}, by  equation $P_{rm}(E,u)\exp[-2(a+h^2/(4a))u]=0.$ Solving this equation, we find $u=u_\pm$ or, in terms of the coordinate $r$,
     \begin{equation}\label{reg-r}r_\pm=\left(\frac{1\pm\sqrt{1-4q_e^2E^8}}{2|q_e|E^4}\right)^{ 1/h} \quad (r_+>1, \ r_-=1/r_+ < 1).\end{equation}
   The circle radii of the turning points
   are
$$\rho_\pm=\frac{1}{E}r^a_\pm.$$

The regions accessible by a particle with  energy $E$ are $u\leq u_-$ ($r\leq r_-$)  or $u\geq u_+$ ($r\geq r_+$). For a given $E <1/\sqrt[4]{2|q_e|}$ there is a potential barrier at $u_-< u < u_+$ ($r_-< r < r_+$), whereas for $E >1/\sqrt[4]{2|q_e|}$ the potential barrier vanishes.

A value of $E$ may be chosen such that $u_-$ and $u_+$ coincide, and the potential barrier vanishes. Equating $u_-$ and $u_+$, we find $u=0$ ($r=1$), $E=1/\sqrt[4]{2|q_e|}\equiv E_0$, and from~\eqref{rad_skor}, \eqref{uskorenie} it follows that $\dot u |_{u=0} = 0$,  $\ddot u |_{u=0} = 0$.
The particle located at $u=u_-=u_+=0$ ($r=1$)  will be at a point of unstable equilibrium.
Actually, the only equilibrium point $u=0$ ($r=1$) along radial timelike geodesics is found by solving the equation $V'_{eff}(u)=0$. The equilibrium is unstable, since $V''_{eff}(0)=-(1/4)h^2/\sqrt[4]{2|q_e|}<0 $, and the effective potential $V_{eff}$ has a maximum at $u=0$.

We consider trajectories of particles with $E <E_0$ moving from the turning points
$u_+$  or $u_-$. It follows from ~\eqref{uskorenie} that  initial accelerations at the
turning points
are
$$ \ddot u_\pm\equiv \ddot u(u_\pm)=\pm(hE^2/4)\sqrt{1-4q_e^2E^8}e^{-2(a+h^{2}/(4a))u_\pm}.$$
The initial acceleration is positive for $u_+$ and negative for $u_-$. A particle starting from rest at $u_+$ moves away from the  singularity at  $u=-\infty$ ($r=0$), and a particle that starts from rest at $u_-$ moves in the opposite direction and falls into this  singularity   in a finite proper time. Namely,
for the latter particle
we have
 from~\eqref{rad_skor}
 \begin{equation*}
 \frac{d\tau}{du}= -\exp[(a+h^{2}/(4a))u]\left(E^2-E_0^2/\sqrt{\cosh(hu)}\right)^{-1/2} ,
\end{equation*}
\begin{equation}\begin{array}{c}\label{tau} \displaystyle
 \tau= \frac{1}{E_0} \int_{-\infty}^{u_-}  \frac{\exp[(a+h^{2}/(4a))u]\sqrt[4]{\cosh(hu_-)\cosh(hu)} }{\left(\sqrt{\cosh(hu)}-\sqrt{\cosh(hu_-)}
 \right)^{1/2}}du
 = \\ \\ \displaystyle
 \frac{1}{\sqrt[4]{2}E_0} \int_0^{r_-}  \frac{\sqrt[4]{(r^{h}_-+r^{-h}_-)(r^{h}+r^{-h})} }{r^{1-a-h^{2}/(4a)}\left(\sqrt{r^{h}+r^{-h}}-\sqrt{r^{h}_-+r^{-h}_-}
 \right)^{1/2}}dr<+\infty,
 \end{array}\end{equation}
 whereas we  expressed the particle energy $E$  in terms of the initial value
$u_\pm \ (r_\pm)$ of the radial coordinate by using Eqs.~\eqref{reg}, \eqref{reg-r}:
$$E=\frac{E_0}{\sqrt[4]{\cosh (hu_\pm)}}=\frac{\sqrt[4]{2}E_0}{\sqrt[4]{r^{h}_\pm+r^{-h}_\pm}}.$$   We also assumed that $\tau =0$ at the starting point $u_-<0 \ (r_-<1)$
and   took into account the series expansion $$\sqrt{r^{h}+r^{-h}}-\sqrt{r^{h}_-+r^{-h}_-}=\frac{h}{2r_-}\frac{r^{h}_-
 -r^{-h}_-}{\sqrt{r^{h}_-+r^{-h}_-}}(r-r_-)+o(r-r_-).$$

Similarly, for the particle that starts from rest at $u_+>0 \ (r_+>1)$ we have
$$\tau = \frac{1}{\sqrt[4]{2}E_0} \int_{r_+}^{+\infty}  \frac{\sqrt[4]{(r^{h}_++r^{-h}_+)(r^{h}+r^{-h})} }{r^{1-a-h^{2}/(4a)}\left(\sqrt{r^{h}+r^{-h}}-\sqrt{r^{h}_++r^{-h}_+}
 \right)^{1/2}}dr=+\infty.$$
 The particle starting from rest at $u_+>0 \ (r_+>1)$ is separated by the potential barrier from the singularity at $u=-\infty \ (r=0)$. It  moves away from the singularity and  reaches  infinity in infinite proper time $\tau=+\infty$.

In order that a radially moving  particle with energy $E<E_0$  could traverse  or touch the wormhole throat $u=u_0 < 0 \  (r=r_0 < 1)$ it must be that
$u_-\geq u_0 \ (r_-\geq r_0)$.
Due to Eqs.~\eqref{u0}, \eqref{reg}, this is equivalent to the following condition:
$$  -\mid\arcosh[(2|q_e|E^4)^{-1}]\mid \geq  - \arctanh(4a/h). $$
From
here we find   a (lower) {\it energy threshold} $E_{thr}$ of  traversability of the wormhole throat for a radially moving massive test particle:
\begin{equation}\label{en_trav}
E\geq E_{thr}\equiv E_0\sqrt[8]{1- 16a^2/h^2} \qquad ( E_0=1/\sqrt[4]{2|q_e|}).\end{equation}

 Radially moving particles starting from rest at $u_->u_0$ with energy $E$ satisfying $E_{thr}<E< E_0$  pass through the throat  and fall to the singularity
  in a finite proper time.
Radial trajectories of massive particles starting from rest at $u_-<u_0$ with energy $E< E_{thr} $ do not  pass through  the wormhole throat.

Massive particles with energy $E= E_{thr} $  start at the turning point
$u_-=u_0$ on
the wormhole throat and fall to the singularity
 in a finite proper time.

 Radially moving particles with energy $E < E_0$ that start from rest at $u_+$
   do not  traverse  the wormhole throat; they move away from the throat and  reach  infinity in  infinite proper time.

In the case $2|q_e|E^4 > 1$, i.~e., when $E>E_0$, the  right-hand side  of~\eqref{rad_skor} is strictly positive and  $ \dot u \neq 0$ for
all real $u$.
A particle with an initial value $u_i$ of the radial coordinate and positive initial radial velocity $\dot u_i =\dot u(u_i)>0 $   moves away from the singularity and  reaches  infinity in
infinite proper time. If $u_i<u_0$, the particle  crosses the wormhole throat in a finite proper time $\tau =\int^{u_0}_{u_i}(d\tau/du)du$.
In the case of negative initial radial velocity $\dot u_i <0 $ the particle moves toward the singularity, crosses the wormhole throat  in a finite proper time $\tau =\int^{u_i}_{u_0} (d\tau/du)du $ (only if  $u_i>u_0$),
and falls to the singularity in a finite proper time $\tau =\int^{u_i}_{-\infty}(d\tau/du)du $ (see Eq.~\eqref{tau}). Thus, for $E>E_0$ the wormhole is traversable for radially moving test particles with non-zero rest masses.

\begin{figure}[h]
\center{\includegraphics[width=0.6\linewidth]{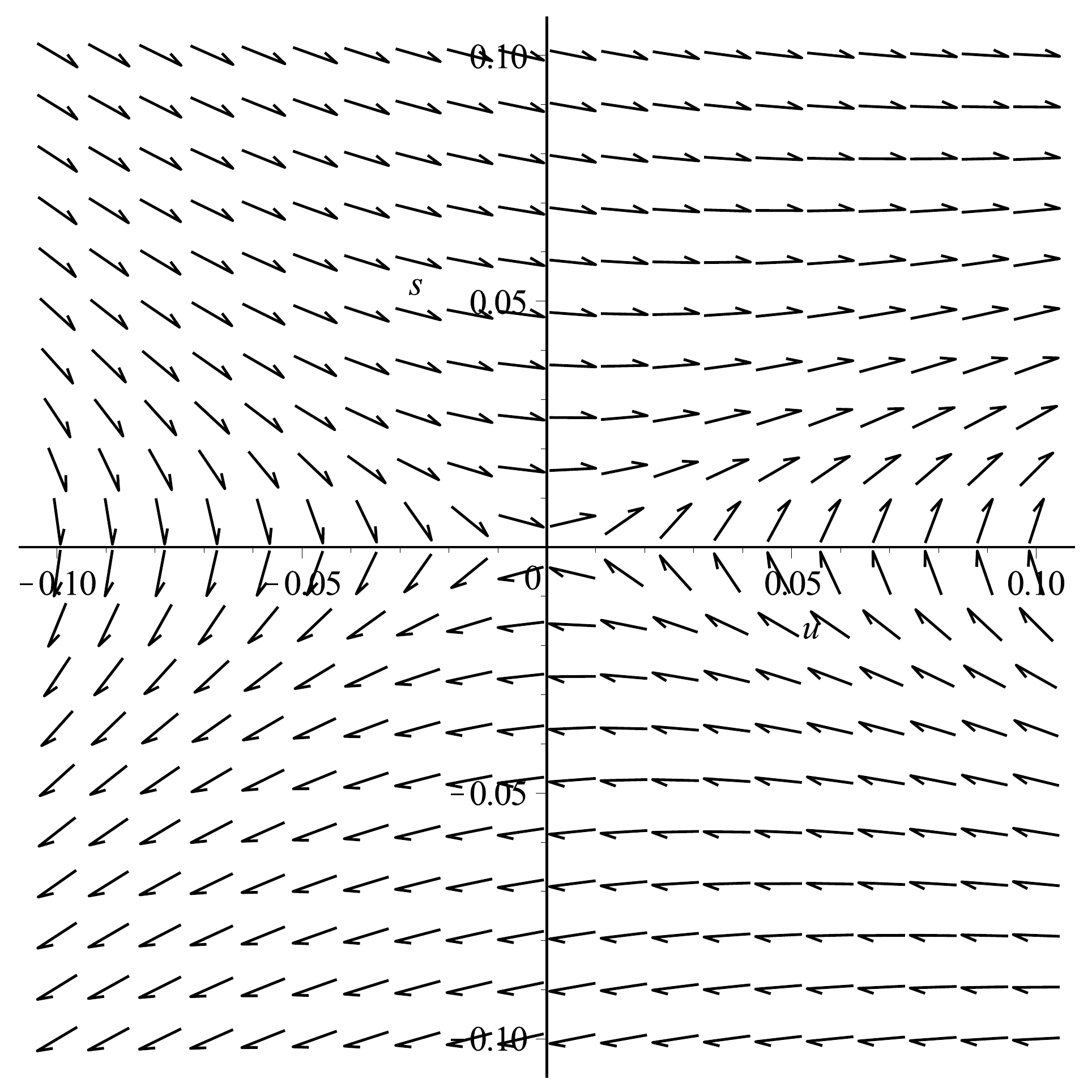}}
\caption{The phase portrait of dynamical system~\eqref{new_form} at fixed $h =1$, $q_e =0.5$, $ a=0.125$. }
  \label{phasePortrait1}
\end{figure}

To obtain information about the possible modes of behavior of particles in the last case
$2|q_e|E^4 = 1$, i.~e., when $E=E_0$, we rewrite Eqs. \eqref{rad_skor}, \eqref{uskorenie} in the form of the dynamical system
\begin{equation}\label{new_form}
 \dot u = s, \qquad  \qquad \dot s =\frac{h}{4}\frac{\tanh(hu)\exp[-2(a+h^2/(4a))u]}{\sqrt{2|q_e|\cosh(hu)}}  -(a+h^2/(4a)) s^2,
\end{equation}
 and study the nature and stability of the corresponding fixed points in the phase plane $(u,s)$. They are given by the equations $\dot u=\dot s=0$ whose solution is the only fixed point $(0,0)$.
The corresponding Jacobian matrix $J$ about the fixed point $(0,0)$ has the form
\begin{equation*}
J(0,0)  =
\begin{pmatrix}
0 \ \ \; 1\\
G_0 \; 0
\end{pmatrix},  \qquad G_0=(1/4)h^2E_0^2.
\end{equation*}
Its eigenvalues $l_\pm =\pm (1/2)hE_0 $ are real and distinct. Therefore, the fixed point $(0,0)$ is  an unstable saddle point.
A typical phase portrait  of dynamical system~\eqref{new_form} at fixed $h =1$, $q_e =0.5$, $ a=0.125$ is presented in Fig.~\ref{phasePortrait1}.

A particle with energy $E_0$ and initial values $u_i<0, \dot u_i<0$ moves toward the singularity, crosses the wormhole throat   (only if  $u_i>u_0$)
and falls to the singularity in a finite proper time
\begin{equation*}\displaystyle
 \tau= \frac{1}{E_0} \int_{-\infty}^{u_i}  \frac{\exp[(a+h^{2}/(4a))u] }{\left(1-\sqrt{\sech(hu)}
 \right)^{1/2}}du
. \end{equation*}
If $u_i<0$, but $ \dot u_i>0$, the particle moves to the hypersurface $u=0$ and reaches it in  infinite proper time
due to the divergence of the integral $\int_{u_i}^{0}(d\tau/du)du$ at $u=0$, which is evident from the series expansion $1-1/\sqrt{\cosh(hu)}=(1/4)h^2u^2+o(u^3)$.

A particle with energy $E_0$ and initial values $u_i>0, \dot u_i>0$    moves away from  the singularity and reaches  infinity in  infinite proper time $\int_{u_i}^{+\infty}(d\tau/du)du$. In the case $u_i>0, \ \dot u_i<0$ the particle moves to the hypersurface $u=0$ and reaches it in  infinite proper time
$\int_{0}^{u_i}(d\tau/du)du$, so the particle does not cross the throat. It turns out that the hypersurface $u=0$ traps out particles with energy $E_0$.



\section{Radial motion of photons.}

From Eq.~\eqref{skorparticle}
for null geodesics $\iota =0$  it follows that
\begin{equation*}
 \dot{u}^2 = \left(E^2e^{-2\gamma}-M_z^2e^{-2\xi}-L^2e^{-2\beta}\right) e^{-2(\gamma+\beta+\xi)}.
\end{equation*}
Since we consider radial geodesics, we put $M_z=L=0$, and the  equation takes  the form
$$  \dot {u} ^ 2=  E^2e^{-2(2\gamma+\beta+\xi)}
$$ or, in view of~\eqref{metr(2)}, the form $$  \dot {u} ^ 2=  E^2e^{-2(a+h^2/(4a))u}.
$$ Integrating, we find
  \begin{equation*}
u(\tau) =
           (a+b)^{-1}\ln[\pm (a+h^2/(4a))E(\tau-\tau_0)].\end{equation*}
The radial coordinate $u$ varies from $-\infty$ to $+\infty$ if $\dot u>0$ (outgoing radial null geodesics) or from $+\infty$ to $-\infty$ if $\dot u<0$ (ingoing radial null geodesics). Each of such geodesics crosses the throat, and the wormhole turns out to be traversable by photons.


\section{Non-radial motion  in the plane $z=\mbox{const} \ (M_z=0)$.}\label{IIC}
\begin{figure}[ht]
\center{\includegraphics[width=0.5\linewidth]{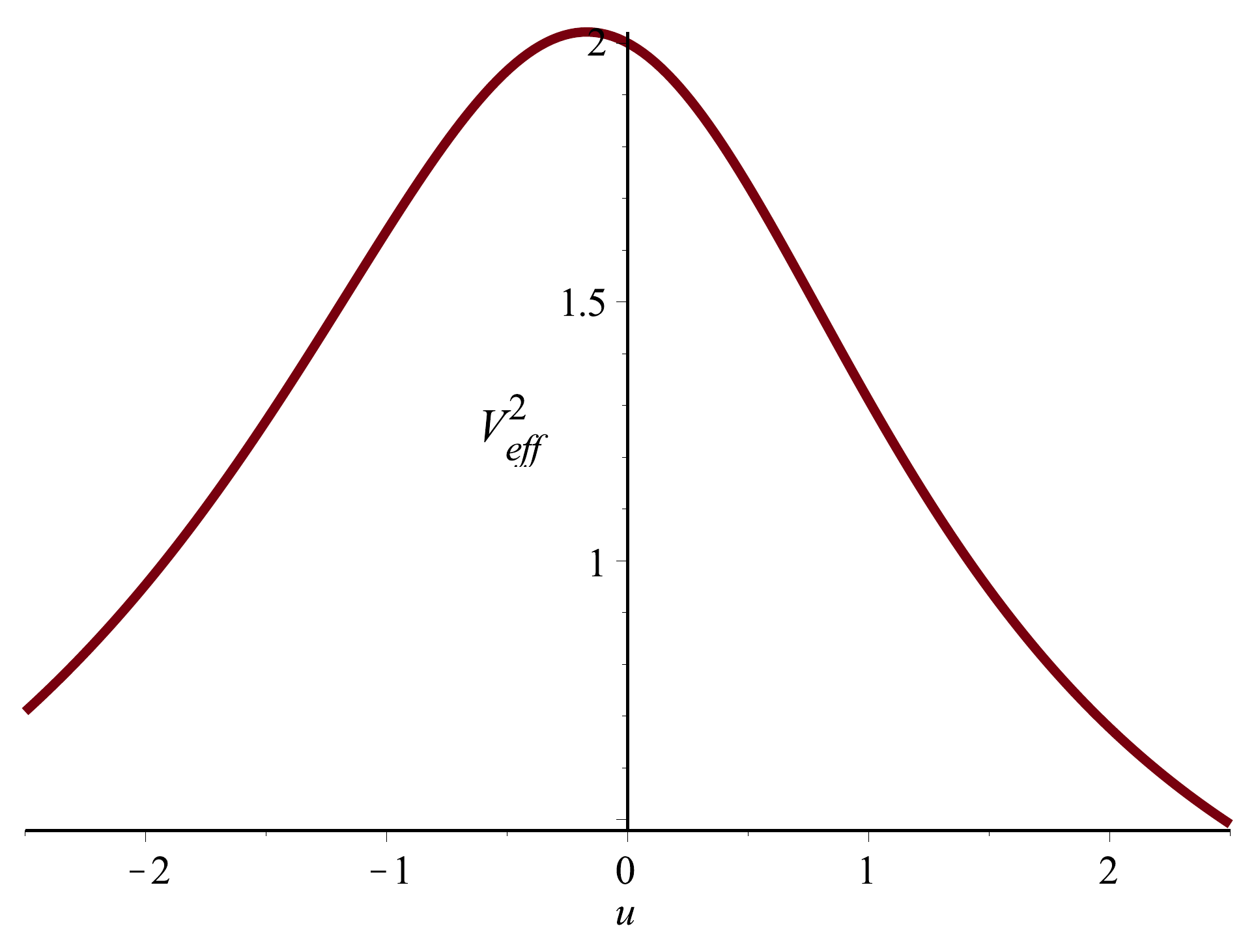}}
\caption{The plot of
$V^2_{eff}(u)$ \eqref{effpot_nonrad} for  timelike geodesics $(\iota=1)$ in the plane $z=\mbox{const} \ (M_z=0)$
at fixed $L = 1$,  $E_0 = 1$, $h = 1$, $a = 1/8$.}
\label{new2_L=1_corr}
\end{figure}

In the case $M_z=0$,  $L\neq 0$ we obtain from~\eqref{iiint1}  $z=\const$,
hence,  particles move along a hypersurface $z=\const$ (``planar orbits").
From Eqs.~\eqref{metr(2)},~\eqref{skorparticle}, \eqref{potencialnaya_energiya}, and \eqref{uskorenie_dlya_metriki_cm1}  we obtain
\begin{equation}\label{dudtau}
 \dot{u}^2 =   \left(E^2
 -E_0^4L^2e^{-2au}\sech(hu)-\iota E_0^2\sqrt{\sech(hu)}\right)  e^{-2(a+b)u},
 \end{equation}
 \begin{equation*}\begin{array}{c}
 \ddot u = \left((1/4)\iota hE_0^2\tanh(hu)\sqrt{\sech(hu)} +\right.\\ \\ \left.
 (1/2)E_0^4L^2\sech(h u) \left(h\tanh(h u)+2a\right)e^{-2au}- (a+b)P(u)\right)e^{-2(a+b)u},
\end{array}\end{equation*}
\begin{equation}\label{effpot_nonrad}
 V_{eff}^2(u) = E_0^2\left(E_0^2L^2e^{-2au} \sech(h u)+\iota \sqrt{\sech(h u)}\right).
 \end{equation}
The plot
of the effective potential $V_{eff}(u)$
for  timelike geodesics $(\iota=1)$ in the plane $z=\mbox{const} \ (M_z=0)$
at fixed
 $L = 1$,  $E_0 = 1$, $h = 1$, $a = 1/8$
is shown  in Fig.~\ref{new2_L=1_corr}.

By considering $u$ as a function of $\phi$ and combining Eqs.~\eqref{iiint1} and~\eqref{dudtau}, we derive the equation
\begin{equation}\label{dudphi}\begin{array}{c}\displaystyle
 \left(\frac{du}{d\phi}\right)^2 = E_0^{-4}L^{-2} \left(E^2
 -E_0^4L^2e^{-2au}\sech(hu)-\iota E_0^2\sqrt{\sech(hu)}\right) \cosh(hu) e^{2(a-b)u} \\ \\
 \equiv E_0^{-4}L^{-2} P(u) \cosh(hu) e^{2(a-b)u}.
\end{array}\end{equation}
Once this equation has been solved, the complete solution can be obtained by direct quadratures of the equations
$$ \frac{d\tau}{d\phi}= (\sqrt{E_0}/L)\sqrt{\cosh(hu)}  e^{2au} \quad \mbox{and} \quad
 \frac{dt}{d\phi}=(EE_0/L) \cosh{(hu)}  e^{2au}. $$

In asymptotic regions, when $u \rightarrow \pm \infty$, we get  $P(u) \rightarrow E^2 - 0$. It follows from Eq.~\eqref{dudphi} that
for the large enough  positive values of the coordinate $u$  we have approximately
\begin{equation*}
 \left|\frac{du}{d\phi}\right| = \frac{1}{\sqrt{2}E_0^{2}|D|} e^{(h/2+a-b)u}
\end{equation*}
or, after integration,
\begin{equation*}
   e^{-(h/2+a-b)u} =  \frac{1}{\sqrt{2}E_0^2|D|}| (h/2+a-b) (\phi - \phi_{+\infty})|, \qquad  \phi_{+\infty}=\const, \end{equation*}
where we introduced an impact parameter $D=L/E$. Since $ 0< a < h/4$ and $h/2+a-b < 0$ (see Eq.~\eqref{cm1}), this implies a spiral character of the particle motion at  infinity $u=+\infty$, i.e., $ \phi \rightarrow \phi_{+\infty}$ \ when \ $ u \rightarrow + \infty$.  By analogy, in the case when $u \rightarrow - \infty$, we get
\begin{equation*}
  e^{-(-h/2+a-b)u} = \frac{1}{\sqrt{2}E_0^2|D|}| (-h/2+a-b) (\phi - \phi_{-\infty})|,   \qquad  \phi_{-\infty}=\const,
 \end{equation*}
 which, because of  the condition $-h/2+a-b < 0$, confirms the radial character of the motion near the singularity, i.e., \ $ \phi \rightarrow \phi_{-\infty}$ \ when \ $ u \rightarrow - \infty$.

 The radii $u_c \ (r_c)$ of circular orbits and the corresponding values of $E$ and $L$ are defined by extrema of $V_{eff}(u)$, namely, the minima correspond to stable orbits, while the maxima correspond to unstable orbits. The joint solution of equations $V_{eff}(u) = E$ and $$V_{eff}'(u)=  -(1/4)V_{eff}^{-1}\left(\iota h\tanh(hu)+2E_0^2L^2 (2a+h\tanh(hu))e^{-2au}\sqrt{\sech(h u)}\right)=0$$
(which are equivalent to equations $P(u)=0$ and $P'(u)=0$)
gives the only radius
$u_c \leq -h^{-1} \arctanh(2a/h)$ ($r_c\leq [(h-2a)/(h+2a)]^{1/(2h)}$), which is equal for null geodesics ($\iota=0$) to
 \begin{equation*}u_c=-h^{-1} \arctanh(2a/h)\quad \quad (r_c= [(h-2a)/(h+2a)]^{1/(2h)}),\end{equation*}
 and  for timelike geodesics ($\iota=1$) it is implicitly defined    by the equation
  \begin{equation*}
2E_0^2L^2 (2a+h\tanh(hu_c))e^{-2au_c}\sqrt{\sech(h u_c)}+h\tanh(hu_c)=0. \end{equation*}
Since $V''_{eff}(u_c)<0$, the circular orbit $u=u_c$ is unstable.
As $V_{eff}$ has no finite minima, there are no stable circular orbits.

 As for the  radial  motion, it can be verified that the fixed point $(u_c,0)$ is an unstable saddle point of the corresponding dynamical system
similar  to system~\eqref{new_form}.

We will show  that, as in the Kerr and Schwarzschild
geometries, we should distinguish between  orbits with  impact parameters greater resp. less than a certain critical value $D_c$ of the impact parameter, which corresponds to the unstable  circular orbit of radius $u_c$ $(r_c)$. For $D^2>D_c^2$ there are two kinds of orbits: orbits of the first kind  arrive from infinity and have pericenter distances greater than $u_c$, whereas orbits of the second kind have apocenter distances less than $u_c$ and terminate at the singularity at $u=-\infty$ $(r=0)$. For $D=D_c$ orbits
of the first and second  kinds merge, and both orbits spiral infinite number of times around the unstable circular orbit  $u=u_c$.
For $D^2<D_c^2$ we have only  orbits
of one kind:  starting at infinity, they cross  the wormhole throat and terminate at the singularity.

As in the radial case, the geometry of the geodesics (photon and massive test particles orbits)
is determined by  the  number and the location of the roots of the equation $P(u)=0$.

It follows from the above that the function $V^2_{eff}(u)$ has the only maximum  at $u=u_c$. Hence,
the function
\beq\label{P(u)_U}
P(u)=L^2\left(D^{-2}-V^2_{eff}(u)/L^2\right)\eeq
has the only minimum at $u_c$. Since $P(u)$ is smooth and its limit equals $L^2/D^2$ when $u$ approaches $+\infty$ or $-\infty$, it  has no more than two real zeros, which we denote by $u_1$ and $u_2>u_1$.
As a result, we have the following three cases described below.

{\it Case} (a):
 $P(u)$ has one zero at $u_c$. From  $P(u_c)=0$  we find the critical value of the impact parameter: $D_c^2\equiv 1/U(u_c).$  Equation \eqref{dudphi}, where $P(u)$ is given by \eqref{P(u)_U} and $D^2=D_c^2$, has solution $u=u_c$, which is the unstable circular orbit at $u_c$.
Expanding this  equation to  second order
in $u-u_c$ and neglecting terms of higher order, we obtain
\begin{equation*}
 \left(\frac{du}{d\phi}\right)^2 =\alpha^2 (u-u_c)^2 \qquad \ (\alpha=\const>0)
\end{equation*}
(where for photon orbits $\alpha^2=(1/2)(h-2a)^{1-b/h}(h+2a)^{b/h}$) or, in the integrated form, $$(i)\ u=u_c+e^{\pm\alpha(\phi- \phi_0)},  \ u>u_c \qquad  (ii) \ u=u_c-e^{\pm\alpha(\phi- \phi_0)} ,\  u<u_c.$$
Eq. (i) shows that $ u \rightarrow u_c+0$ when $\phi \rightarrow \mp \infty$, i.e., an orbit asymptotically approaches the circle at $u_c$  spiraling around it in clockwise or in counter-clockwise directions an infinite number of times.
This is an orbit of the first kind.
In the interior of the circular orbit $u=u_c$, one obtains orbits of the second kind. Derived from Eq.~(ii), they   unwind from the circular orbit and fall to the singularity.

For the null geodesics (the photon orbits)  the function $P(u)$ (see Eq.~\eqref{dudtau} and Eq.~\eqref{dudphi}) becomes
$$
P(u) \equiv L^2\left( D^{-2} - E_0^4 e^{-2au} \sech(h u)\right),$$
\begin{equation}\label{dudphi_null}
 \left(\frac{du}{d\phi}\right)^2 = E_0^{-4}\left( D^{-2} - E_0^4 e^{-2au} \sech(h u)\right) \cosh(hu) e^{2(a-b)u},
 \end{equation}
or, in the coordinate $r$,
\begin{equation*}
 \left(\frac{dr}{d\phi}\right)^2 = \frac{1}{2}D^{-2}E_0^{-4}r^{2(a-b+1)}\left(r^h+r^{-h}\right) - r^{2(1-b)} .
\end{equation*}

\begin{figure}[h]
\center{\includegraphics[width=0.5\linewidth]{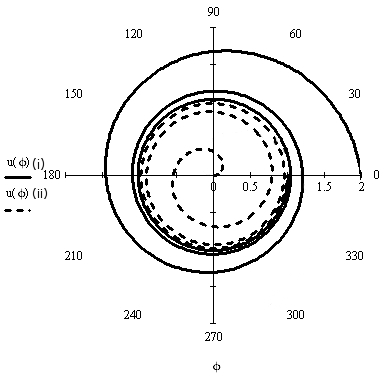}}
\caption{The photon orbits $u=u(\phi)$ of the first kind (the solid line) and the second kind (the dashed line)  in the plane $z=\mbox{const}$ for $D^2=D^2_c$.}
\label{polar2_3}
\end{figure}

The photon orbits $u=u(\phi)$ of the first kind are shown in Fig.~\ref{polar2_3}
by the solid line and those of the second kind  by the dashed line.  Since $u_{th}\equiv - h^{-1} \arctanh(4a/h)  < u_c \equiv - h^{-1} \arctanh(2a/h)$, the photon orbits of the first kind cannot cross the throat, but the photon orbits of the second kind can traverse the throat.

{\it Case} (b):
 $P(u)$ has no zeros, i.~e., $P(u)>0$ for all $u\in \mathbb{R}$. We use Eq.~\eqref{P(u)_U} to rewrite Eq.~\eqref{dudphi} in the form
 \begin{equation}\label{dudphiD}
 \left(\frac{du}{d\phi}\right)^2 = E_0^{-4}D_c^{-2} \left(D_c^2 /D^{2} -U(u)/U(u_c)\right)e^{2(a-b)u}\cosh(hu).
\end{equation}
Taking into account that $U(u)$ has maximum $U(u_c)$ at $u_c$ and, hence, $0<U(u)/U(u_c)\leq 1$ for  all  $u$, we see that $D_c^2 /D^{2}>1$, i.~e., $D^2 <D_c^{2}$. The radial coordinate varies from $-\infty$ to $+\infty$, and $du/d\phi$ vanishes nowhere. Thus, particles starting at infinity cross the wormhole throat and fall to the singularity, so  the wormhole is  traversable.

{\it Case} (c):
 $P(u)$ has two real zeros $u_1$, $u_2$ and $u_1<u_c<u_2$. This is possible only if $D^2 >D_c^{2}$.
 For every value of the impact parameter $D$ there exist two distinct orbits confined to the intervals $u\leq u_1$ and $u\geq u_2$, respectively. They are derived from the  expansion of Eq.~\eqref{dudphiD} in the neighbourhoods of points $u_1$ and $u_2$, namely, $$({du}/{d \phi})^2 = \pm \alpha^2 (u-u_{1,2}) + o(u-u_{1,2}),\quad\alpha =\const>0.$$ Neglecting higher-order terms and  integrating, we get  orbits of the first kind
 $ u=u_2 + \alpha (\phi-\phi_0)^2,\ u\geq u_2$, arriving from infinity and having pericenter distance $u_2$ and the orbits of the second kind $ u=u_1 - \alpha (\phi-\phi_0)^2,\ u\leq u_1$, having the apocenter distance $u_1$ and terminating at a singularity at $u=-\infty$ $(r=0)$.

\section{Non-radial motion in the plane $\phi=\mbox{const} \ (L=0)$.}
\begin{figure}[ht]
\center{\includegraphics[width=0.5\linewidth]{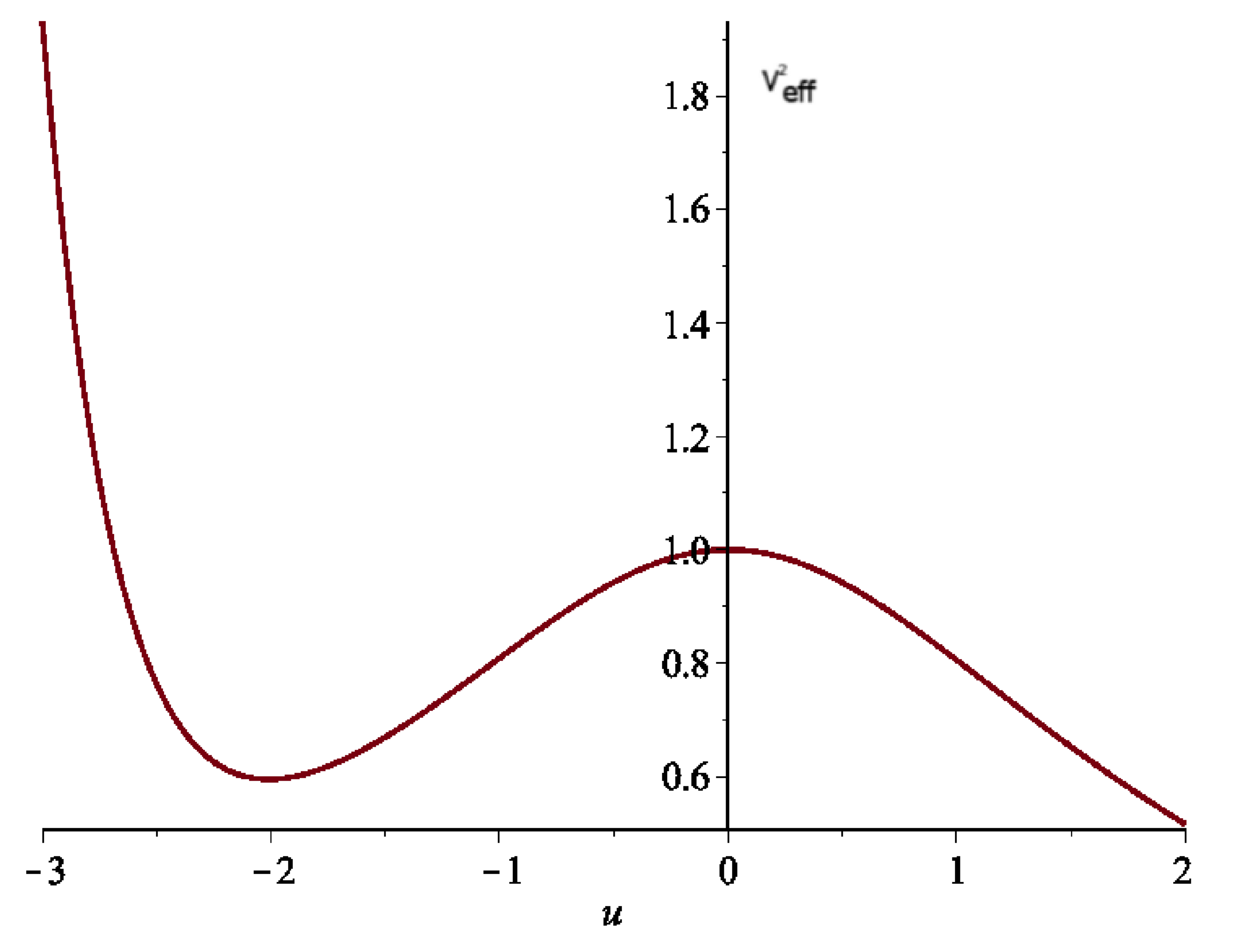}}
\caption{The plot of
$V^2_{eff}(u)$ \eqref{sqV_eff_M=0} for  timelike geodesics $(\iota=1)$ in the  plane $\phi=\mbox{const} \ (L=0)$
at fixed $M_z = 10^{-2}$,  $E_0 = 1$, $h = 1$, $a = 1/8$, $b=2$.
}
 \label{sqV_eff_M=0.pdf}
\end{figure}

In the case $M_z\neq 0$, $L=0$  it follows from~\eqref{iiint1} that $\phi=\const$,
so  particles move along a hypersurface (``plane")\  $\phi=\const$.
As in the preceding section  we
  have from Eqs.~\eqref{metr(2)},~\eqref{skorparticle}, \eqref{potencialnaya_energiya}, and
\eqref{uskorenie_dlya_metriki_cm1}
\begin{equation}\label{p2}
\left({\frac{du}{d\tau}}\right)^2 = \left(E^2-\iota E_0^2 \sqrt{\sech(hu)}-E_0^4M_z^2e^{-2bu}\sech(hu)\right) e^{-2(a+b)u}\equiv
{\widetilde P}(u)  e^{-2(a+b)u}
\end{equation}
or
\begin{equation*}
\left(\frac{du}{d\tau}\right)^2 =\left(E^2 - V_{eff}^2(u)\right)   e^{-2(a+b)u},\end{equation*}
where

\beq\label{sqV_eff_M=0}
V^2_{eff}(u)=E^2_0\left(\iota  \sqrt{\sech(hu)}+E_0^2M_z^2 e^{-2bu}\sech(hu)\right) \eeq
is the squared effective potential shown for timelike geodesics $(\iota=1)$ in Fig.~\ref{sqV_eff_M=0.pdf}.

 Eqs.~\eqref{iiint1} and \eqref{p2} give
\begin{equation}\label{dudz}
 \left(\frac{du}{dz}\right)^2 = E_0^{-4} M_z^{-2}  {\widetilde P}(u) \cosh(hu)  e^{2(b-a)u}.
\end{equation}
The complete solution of the last equation is obtained by  quadratures of the equations
$$ \frac{d\tau}{dz}= \frac{1}{E_0^2M_z}\sqrt{\cosh(hu)}  e^{2bu} \quad \mbox{and} \quad
 \frac{dt}{dz}=\frac{E}{E_0^4M_z}\cosh{(hu)}  e^{2bu}. $$

Eqs.~\eqref{p2} and \eqref{dudz}
for large values of $u$ ($u\rightarrow + \infty$) imply
\begin{equation*}
 \left(\frac{du}{dz}\right)^2 =\frac{1}{2} E_0^{-4} H^{-2}  e^{(2b-2a + h)u}, \qquad H \equiv M_z/E,
\end{equation*}
\begin{equation*}
 \left(\frac{du}{d\tau}\right)^2 = E^2  e^{-2(a+b)u}
\end{equation*}
or, in the integrated form,
\begin{equation*}
  e^{-(h/2+b-a)u} = - \frac{1}{\sqrt{2}E_0^{2}|H|} (h/2+b-a) (z - z_{+\infty}),
\end{equation*}

\begin{equation*}
  e^{(a+b)u} =  E (a+b)(\tau - \tau_0).
\end{equation*}
Since $h/2+b-a>0$, it follows that  $z \rightarrow z_{+\infty} - 0 \neq +\infty$
 and $\tau \rightarrow +\infty  $ when  $u \rightarrow + \infty$, which means that
 particles reach  (radial) infinity in   infinite proper time.

The extrema  of  the effective potential $V_{eff}(u)$ define the radii $u_c$ of  orbits: the minima (resp., the maxima) correspond to stable (resp., unstable) orbits.
The joint solution of the equations $V_{eff}(u)=E$ (or ${\widetilde P}(u) =0$) and   $V_{eff}'(u)=0$ (or ${\widetilde P}'(u)=0$) gives
a  raduis $u_c$, that determines a fixed point of Eq.~\eqref{p2}.
Calculating the first derivative of $V_{eff}(u)$,
we get
\begin{equation*}
 V_{eff}'(u)=  -(1/4)V_{eff}^{-1}\left(\iota h\tanh(hu)\sqrt{\sech(h u)}+2E_0^2M_z^2 (2b+h\tanh(hu))e^{-2bu}\sqrt{\sech(h u)}\right).
\end{equation*}
For null orbits ($\iota =0$) the equation $V_{eff}'(u) = 0$ has no real roots, since $2b/h >1$,
 and the function ${\widetilde P}(u)=E^2-V_{eff}^2(u)$ has the only  zero at $u_1$.
Consequently, all null orbits have a pericenter distance  $u_1$ and terminate at radial infinity.

\begin{figure}[ht]
\center{\includegraphics[width=0.5\linewidth]{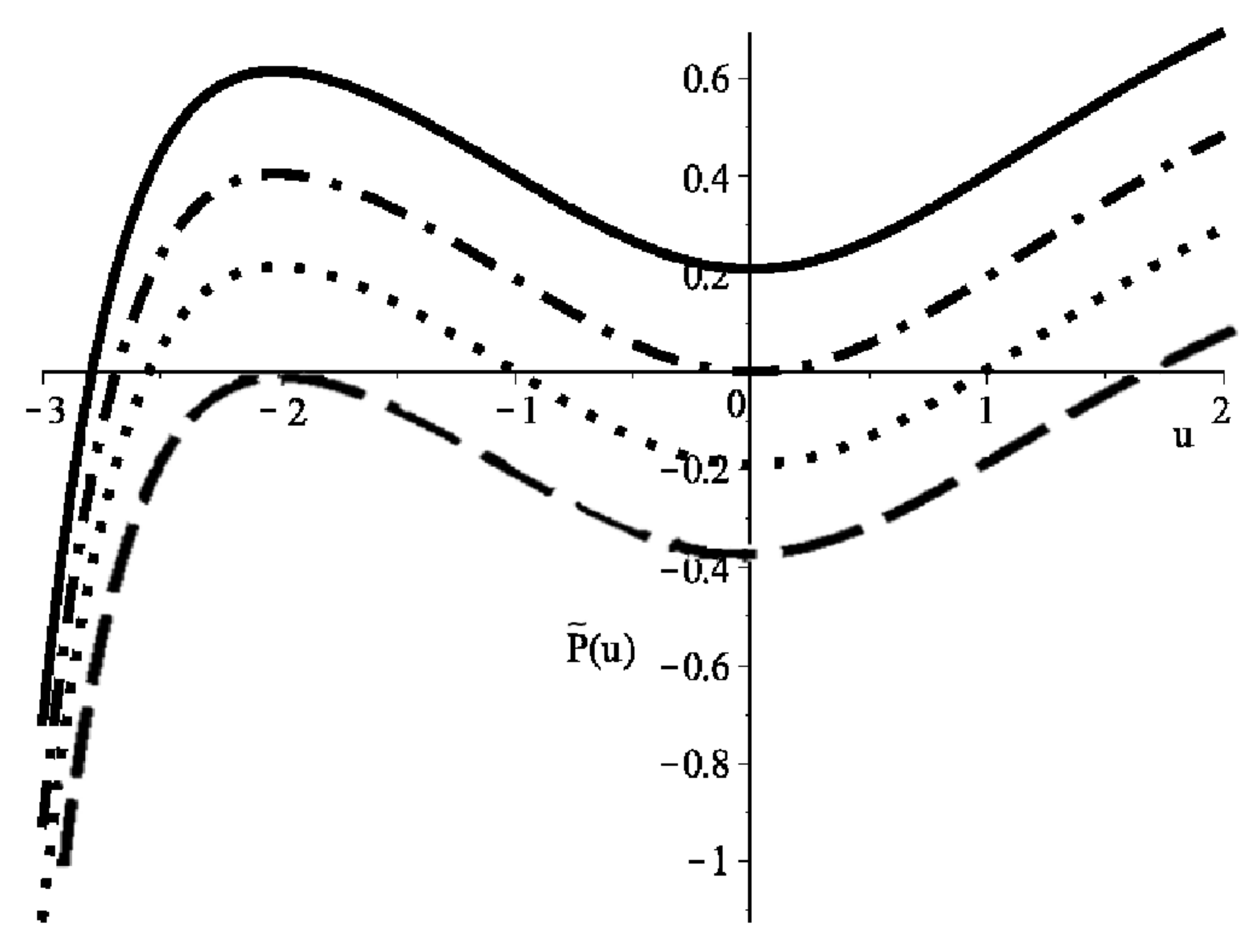}}
\caption{Disposition of zeros of the function ${\widetilde P}(u)$ in the case of timelike geodesics at fixed $M_z = 10^{-2}$,  $E_0 = 1$, $h = 1$, $a = 1/8$, $b=2$, $E=1.1$ (the solid line),   $E=1.0$ (the dash-dotted line),  $E=0.9$ (the dotted line), $E=0.77$ (the dashed line).}
\label{ris1}
\end{figure}

For  timelike orbits ($\iota=1$) the
equation $V_{eff}'(u) = 0$, which can be written in the form
\begin{equation*}
\tanh(h u) = - \frac{2 [b/h]}{
 (2M_z^2 E_0^2 )^{-1}
 \exp(2bu)\sqrt{
\cosh(h u)} + 1},
\end{equation*}
can allow no more than  two real roots.
 Therefore, the equation $V_{eff}(u) = E$ (or ${\widetilde P}(u) =0$) can allow no more than three real roots.
Since
 $\lim_{u \rightarrow + \infty} {\widetilde P}(u) = E^2 >0$ and  $ \lim_{u \rightarrow - \infty} {\widetilde P}(u)  = -\infty,$
the equation  ${\widetilde P}(u)=0$ must always have at least one real root.
Thus, we have to distinguish between the following four cases (see Fig.~\ref{ris1}).

\textit{Case} ($a$):  ${\widetilde P}(u)$ has three distinct real zeros at $u_1, u_2,u_3:$ $u_1<u_2 <u_3$  (the dotted line in Fig.~\ref{ris1}). There exist two kinds of orbits. An orbit of the first kind oscillates between two values of $u$ with $u_1$ being a pericenter and
$u_2$ an apocenter. An orbit of the second kind arrives from infinity and has the pericenter distance
$u_3$.

  \textit{Case} ($b$): ${\widetilde P}(u)$ has two distinct real zeros at $u_1, u_2:$ $u_1<u_2$, ${\widetilde P}'(u_1)\neq 0$, ${\widetilde P}'(u_2) = 0$. Then
  ${\widetilde P}(u)$ has a minimum and $V_{eff}(u)$ has a maximum at $u_2$  (the dash-dot line in Fig.~\ref{ris1}). We have the unstable orbit  $u=u_2$. The orbit of the first kind starts at the pericenter distance $u_1$ and approaches the   orbit $u=u_2$ asymptotically in  infinite proper time. The orbit of the second kind arrives from infinity and has the pericenter distance
  $u_2$.

  \textit{Case} ($c$): ${\widetilde P}(u)$ has two distinct real zeros at $u_1, u_2:$ $u_1<u_2$, ${\widetilde P}'(u_1)= 0$, ${\widetilde P}'(u_2) \neq 0$. Then
  ${\widetilde P}(u)$ has a maximum and $V_{eff}(u)$ has a minimum at $u_1$  (the dashed line in Fig.~\ref{ris1}). We have a stable  orbit  $u=u_1$ and an orbit that arrives from infinity and has the pericenter distance  $u_2$.

  \textit{Case} ($d$): ${\widetilde P}(u)$ has one zero at $u_1$ (the solid line in Fig.~\ref{ris1}).
 Orbits arrive from infinity and have the pericenter distance  $u_1.$

\section{Conclusion}

We studied the  radial and non-radial motion of  test particles for cylindrically symmetric wormholes of  type $\rm WhCR^e$  found earlier \cite{AmChum2013} in the 6-dimensional reduced Kaluza--Klein theory with the Abelian gauge field and two dilaton fields. We  showed that space--time~\eqref{cm1} has a regular wormhole region and has a physical singularity at $u = -\infty$ (for $a,b >0$).
 The motion of test particles in the $\phi$ and $z$-direction is governed by two conserved conjugate momenta $L$ and $M$, and the radial motion is described by an effective potential as in Fig.~\ref{V_{eff}(u)}. This potential is repulsive away from the wormhole throat region. This difference from the typically attractive nature of spherically symmetric wormholes is related to the collapse of the latter, which allows a particle at the throat to move to a smaller circle radius, whereas in our example the  size of the throat does not change. Thus,
 for a  radially moving particle with  energy $E <1/\sqrt[4]{2|q_e|}$ there is a potential barrier at $u_-< u < u_+$.
 Equating $u_-$ and $u_+$ gives $u=0$, $E=1/\sqrt[4]{2|q_e|}\equiv E_0$, so the particle located at $u=u_-=u_+=0$  will be at a point of (unstable) equilibrium. The energy $E_0$ divides motions of qualitatively different behavior.
 The particle with energy $E<E_0$ starting from rest at $u_+>0$ is separated by the potential barrier from the singularity at $u=-\infty$. It  moves away from the singularity and  reaches  infinity in infinite proper time $\tau=+\infty$.
 In order that a radially moving  particle with energy $E<E_0$  could traverse  or touch the wormhole throat $u=u_0 < 0 $ it must have $u_-\geq u_0 $.
 From
 this follows the existence of  a (lower) {energy threshold} $E_{thr}\equiv E_0\sqrt[8]{1- 16a^2/h^2}$ of  traversability of the wormhole throat for a radially moving massive test particle.
  Particles starting from rest at $u_->u_0$ with energy $E$ satisfying $E_{thr}<E< E_0$  pass through the throat  and fall to the singularity
   in a finite proper time.

Radial trajectories of massive particles starting from rest at $u_-<u_0$ with energy $E< E_{thr} $ do not  pass through  the wormhole throat.
Massive particles with energy $E= E_{thr} $  start at the point of rest $u_-=u_0$ on
the wormhole throat and fall toward the singularity
 in a finite proper time.
 Radially moving particles with energy $E < E_0$ that start from rest at $u_+$
   do not  traverse  the wormhole throat; they move away from the throat and  reach  infinity in infinite proper time.
 For $E>E_0$ the wormhole is traversable for radially moving test particles with non-zero rest masses.

A particle with energy $E_0$ and initial values $u_i<0, \dot u_i<0$  of radial coordinate and velocity  moves toward the singularity, crosses the wormhole throat   (only if  $u_i>u_0$)
and falls to the singularity in a finite proper time.
If $u_i<0$, but $ \dot u_i>0$, the particle moves to the hypersurface $u=0$ and reaches it in infinite proper time. When $u_i>0, \dot u_i>0$
 a particle  moves away from  the singularity and reaches  infinity in infinite proper time. In the case $u_i>0, \ \dot u_i<0$ the particle moves toward the hypersurface $u=0$ and reaches it in infinite proper time, so the particle does not cross the throat (the hypersurface $u=0$ traps out particles with energy $E_0$).
 Photons, being scale (conformally) invariant, cannot have a critical energy $E_0$, and in fact we found that
 each radial photon trajectoriy crosses the throat.

In the case of non-radial motion the effective potential has a ``centrifugal potential" contribution due to the conserved canonical momenta $M$ and $L$ of the $z$- and $\phi$-motion. As shown by Eq (19), this contribution is always positive, and therefore increases the potential barrier compared to that for pure radial motion. For motion in the hypersurface $z =$ const (``planar orbits", $M=0$, $L\neq 0$) the ability of a particle to overcome this barrier can be stated in terms of its impact parameter $D$.  As in the Kerr and Schwarzschild
 geometries we should then distinguish,   \cite{chan}, between orbits with  impact parameters greater and/or less than a certain critical value $D_c$ of the impact parameter, which corresponds to the unstable  circular orbit of radius $u_c$ $(r_c)$. For $D^2>D_c^2$ there are two kinds of orbits: orbits of the first kind  arrive from infinity ($u=+\infty$) and have pericenter distances greater than $u_c$, whereas  orbits of the second kind have apocenter distances less than $u_c$ and terminate at singularity at $u=-\infty$. For $D=D_c$ orbits
 of the first and second  kinds  spiral infinite number of times on the unstable circular orbit  $u=u_c$.
 For $D^2<D_c^2$ we have only  orbits
 of one kind:  starting at infinity, they cross  the wormhole throat and terminate at the singularity. Thus, to penetrtate the wormhole, improperly (non-radially) aimed planar orbits need more energy and time to traverse the wormhole, but they still do not avoid the central singularity.

 In the case of non-radial motion in the plane  $\phi = $ const  the effect of the corresponding ``centrifugal potential" due to $M$ is similar to that due to $L$ except that $V_{eff}$ always diverges at $u = -\infty$, shielding the singularity: all orbits have a finite pericenter distance. All null orbits have a pericenter distance  $u_1$ and terminate at radial infinity. For massive particles we have to distinguish between the  four cases ($a,b,c,d$). In case  ($a$)  there exist two kinds of orbits: an orbit of the first kind oscillates between two values of $u$ with $u_1$ being a pericenter and
 $u_2$ an apocenter, an orbit of the second kind arrives from infinity and has the pericenter distance $u_3>u_2>u_1$. In case
 ($b$) we have the unstable orbit  $u=u_2$; the orbit of the first kind starts at the pericenter distance $u_1<u_2$ and approaches the   orbit $u=u_2$ asymptotically in the infinite proper time, and the orbit of the second kind arrives from infinity and has the pericenter distance
   $u_2$. In case ($c$) we have the stable  orbit at $u=u_1$ and an orbit that arrives from infinity and has the pericenter distance  $u_2>u_1$.
 In case ($d$)
  orbits arrive from infinity and have the pericenter distance  $u_1.$
This type of orbit allows traversing the wormhole from $V_+$, spending a finite time on the "other side" $V_-$ without encountering the singularity, and reemerging in the original space $V_+$.

\end{document}